\documentclass[structabstract]{aa}  
\usepackage{graphicx}
\usepackage{ulem}
\def\arcsec{$^{\prime\prime}$}

\begin{document}

   \title{Polarization of cluster radio halos with upcoming radio interferometers}
  
   \subtitle{}

   \author{F. Govoni \inst{1},
           M. Murgia \inst{1},
           H. Xu \inst{2,3},
           H. Li \inst{3},
           M.L. Norman \inst{2},
           L. Feretti \inst{4},
           G. Giovannini \inst{4,5},
           V. Vacca  \inst{4,5}
           }

  \institute{INAF - Osservatorio Astronomico di Cagliari,
             Strada 54, Loc. Poggio dei Pini, 09012 Capoterra (Ca), Italy
          \and 
             Center for Astrophysics and Space Science, 
             University of California at San Diego, La Jolla (CA), USA
          \and 
             Theoretical Division, Los Alamos National Laboratory, Los
             Alamos (NM), USA
          \and 
             INAF - Istituto di Radioastronomia, Via P.Gobetti 101, 
             40129 Bologna, Italy 
          \and 
             Dipartimento di Fisica e Astronomia, Universit\`a degli Studi di Bologna, 
             Via Ranzani 1, 40127 Bologna, Italy
             }

   \date{Received September 15, 1996; accepted March 16, 1997}

  \abstract
  {Synchrotron radio halos at the center of merging galaxy clusters 
   provide the most spectacular and direct evidence  
   of the presence of relativistic particles and magnetic fields associated 
   with the intracluster medium. The study of polarized emission from
   radio halos has been shown to be extremely important to 
   constrain the properties of intracluster magnetic fields. 
   However, detecting this polarized signal is a 
   very hard task with the current radio facilities.}
   {We investigate whether future radio observatories, 
   such as the Square Kilometer Array (SKA), its precursors 
   and its pathfinders, will be able 
   to detect the polarized emission 
   of radio halos in galaxy clusters.}
     {On the basis of cosmological magnetohydrodynamical simulations 
      with initial magnetic fields injected by active 
      galactic nuclei, 
      we predict the expected radio halo polarized signal at 1.4 GHz. 
     We compare these expectations with the limits of current radio 
     facilities and 
     explore the potential of the forthcoming radio interferometers
     to investigate intracluster magnetic fields through 
     the detection of polarized emission from radio halos.
    }
  {The resolution and sensitivity values that are expected to be obtained
   in future sky surveys performed at 1.4 GHz using the SKA precursors 
   and pathfinders (like APERTIF and ASKAP) are very promising for the 
   detection of the polarized emission of the most powerful 
   ($L_{1.4GHz}>10^{25}$ Watt/Hz) radio halos. Furthermore, 
   the JVLA have the potential to already detect polarized emission 
   from strong radio halos, at a relatively low resolution.
   However, the possibility of detecting the polarized signal in 
   fainter radio halos ($L_{1.4GHz}\simeq10^{24}$ Watt/Hz) at high resolution
   requires a sensitivity reachable only with SKA.}
   {}


   \titlerunning{Polarization of cluster radio halos with forthcoming radio interferometers}
   \authorrunning{F. Govoni et al.}
   \maketitle
%

\section{Introduction}

According to the hierarchical scenario of structure formation, galaxy
groups and sub-clusters merge together into more massive clusters
releasing amounts of gravitational energy as high as $10^{64}$ erg.
Shocks and turbulence associated with cluster merger 
events are thought to heat thermal gas, to accelerate relativistic particles,
and to compress and amplify magnetic fields in the intracluster medium 
(e.g. Roettiger et al. 1999). 

Sensitive radio observations have revealed diffuse emission from the
central regions of many merging galaxy clusters.
These radio sources, which extend over volumes
of $\sim$1 Mpc$^3$, are called radio halos. They are diffuse,
low-surface-brightness ($\simeq$ 1$\mu$Jy~arcsec$^{-2}$ at 1.4 GHz),
and steep-spectrum\footnote{$S(\nu)\propto \nu^{- \alpha}$, with
$\alpha$=spectral index} ($\alpha>1$) synchrotron sources with
no obvious optical counterparts.
Radio halos demonstrate the existence of relativistic electrons and  
magnetic fields spread in the intracluster medium.
Most of what we know about intracluster magnetic fields derives from
studies of these large-scale cluster diffuse radio sources 
and from Faraday rotation measures 
of polarized radio galaxies located inside or behind galaxy clusters
(see Feretti et al. 2012 for a recent review).

Under the assumption that radio halos are on the minimum energy
condition, and hence their magnetic energy density is comparable
to the energy density in relativistic electrons, it is found 
that the volume-averaged magnetic field is of the order of 
$\sim$ 0.1$-$1 $\mu$G and the total magnetic energy is as high 
as 10$^{61}$ erg.
By comparing the rotation measures observed in seven radio sources 
which are projected or belonging to the Coma cluster with simulations, 
Bonafede et al. (2010) found a cluster magnetic field of a few $\mu$G. 
Similar results have been obtained
by Bonafede et al. (2011a) by studying the trend of the 
fractional polarization of radio sources in a large sample of 
clusters.

Important information on the intracluster 
magnetic field structure can be derived from the analysis 
of detailed radio halo images. The total intensity and the
polarization intensity radio halo surface brightness fluctuations
are strictly related to the magnetic field power 
spectrum (Tribble 1991, Murgia et al. 2004). 
For example, radio halos with a lack of polarization and a 
smooth, regular surface brightness may indicate that the cluster 
magnetic field is tangled on scales smaller than the resolution 
of the radio images, while a disturbed radio morphology and the 
presence of polarization patches could be related to a magnetic field 
ordered on scales larger than the observing beam. 
Thus, observations of radio halos have been used to study 
the structure of the cluster-wide magnetic fields by comparing 
observations with mock halos from turbulent magnetic fields by 
construction (Murgia et al. 2004, Govoni et al. 2006, Vacca et al. 2010). 
In a study of the magnetic field power spectrum in the galaxy cluster A665, 
which contains a Mpc-scale radio halo, the modeling performed
by Vacca et al. (2010)
suggests that although radio halos are usually found to be
unpolarized, they can still be intrinsically polarized; however, 
detecting this polarized signal is a very hard task with the current 
radio interferometers because of their faintness.  
Polarized emission from radio halos have been observed only in bright 
filaments of the clusters A2255 (Govoni et al. 2005, Pizzo et al. 2011) 
and MACS J0717+3745 (Bonafede et al. 2009).

Cosmological simulations have been playing an important part 
in studying the intracluster magnetic field evolution
of galaxy clusters (e.g. Dolag et al. 1999, 2002, 2005, Br{\"u}ggen et al. 2005,
Dubois \& Teyssier 2008, Xu et al. 2009, Bonafede et al. 2011b). 
Although the existence of cluster-wide magnetic fields is now well-established, 
their origin, which is ultimately important for understanding the 
evolution of the intracluster medium during the course of cluster formation, 
is still unclear (Widrow 2002, Dolag et al. 2008).
Magnetohydrodynamical simulations of cluster formation have been performed 
with different initial magnetic fields, which have included 
random or uniform fields from high redshifts (Dolag et al. 2002, Dubois \& Teyssier 2008, Dubois e al. 2009), or from the outflows of normal galaxies 
(Donnert et al. 2009) or from active 
galaxies (Xu et al. 2009, 2010, 2011, 2012).  
The cluster magnetic fields of all these simulations are roughly in agreement 
 at low redshifts. These simulations predict $\mu$G level magnetic field
strengths in the cluster centers and a decrease of the 
magnetic field strengths with radius in the outer regions. 
These simulations provide a variety of merger 
configurations for evolving galaxy clusters, which can be 
used to interpret observed radio and X-ray features of particular 
systems (e.g. Donnert et al. 2010, Bonafede et al. 2012).

Recently, Xu et al. (2012) used cosmological 
magnetohydrodynamical cluster simulations
with initial magnetic fields injected by active galactic nuclei
to generate synthetic Faraday rotation measures and total intensity 
radio-halo images. 
The resulting Faraday rotation measures are consistent 
with the observed values of radio sources in clusters.
Additionally, by giving a reasonable energy spectrum for the synchrotron 
electrons and by assuming the energy equipartition between magnetic fields 
and non-thermal electrons, the resulting radio halos 
have global properties in line with the observations.
The synthetic Faraday rotation measure and total intensity
radio halo images presented by Xu et al. (2012) show 
that the cluster wide magnetic fields that originate 
from active galaxies and are then amplified by the intracluster turbulence
match on the first order with the magnetic field strength and structure 
which are observed in galaxy clusters.

In this work, we make a step forward by predicting the expected 
radio halo polarized signal at 1.4 GHz on the basis of the cosmological
magnetohydrodynamical simulations by Xu et al. (2012).
These simulations provide radio halo images that are compatible 
with data, as far as the total intensity signal is considered. 
We now investigate using the software FARADAY (Murgia et al. 2004) 
the synthetic radio halo polarization, and we compare
these expectations with the current observational upper limits.
Furthermore, we explore the possibility of detecting the polarized signal
with the next generation of radio interferometers.
Our base of knowledge on cluster magnetic fields
will be greatly enhanced by radio observatories, such as the 
SKA\footnote{http://www.skatelescope.org/}, its precursors, and 
its pathfinders which will provide the necessary sensitivity 
to study the details of total intensity and polarization of radio halos 
at GHz frequencies in a large number of galaxy clusters. 

This paper is organized as follows.
In Sect. 2, we present the expected polarized signal of mock radio halos.
In Sect. 3, we show that this faint polarized emission is undetectable 
if observed with the comparatively shallow sensitivity and low resolution 
of current radio interferometers.
In Sect. 4, we use our magnetohydrodynamical simulations 
to explore the potential of the forthcoming large radio telescopes 
to constrain cluster magnetic fields through 
the detection of polarized emission from cluster radio halos.
Finally, we draw the conclusions in Sect. 5. 

The intrinsic parameters quoted in this paper are computed for a 
$\Lambda$CDM cosmology with $H_0$ = 73 km s$^{-1}$Mpc$^{-1}$,
$\Omega_m$ = 0.27, and $\Omega_{\Lambda}$ = 0.73.

\section{Simulated radio halo polarized emission}

\begin{table*}
\caption{Adopted parameters for the relativistic electron distribution 
(see Xu et al. 2012 for details).}  
\label{electrons}      
\centering          
\begin{tabular}{c l l}    
\hline\hline       
    Parameter     & Value                    & Description          \\ 
\hline   
  $\gamma_{min}$   & 300            & Minimum relativistic electron Lorentz factor \\
  $\gamma_{max}$   & 1.5$\times$10$^{4}$ & Maximum relativistic electron Lorentz factor \\
  $\delta$        &    3.6         & Power-law index of the energy spectrum \\
                  &                & of the relativistic electrons \\
  $K_0$           & Adjusted to guarantee & Electron spectrum normalization \\ 
                  & $u_{el}=u_B$ at each point of              &  \\ 
                  & the computational grid                   &  \\ 
\hline                                 
\end{tabular}
\end{table*}

\begin{figure*}
\centering
\includegraphics[width=12 cm]{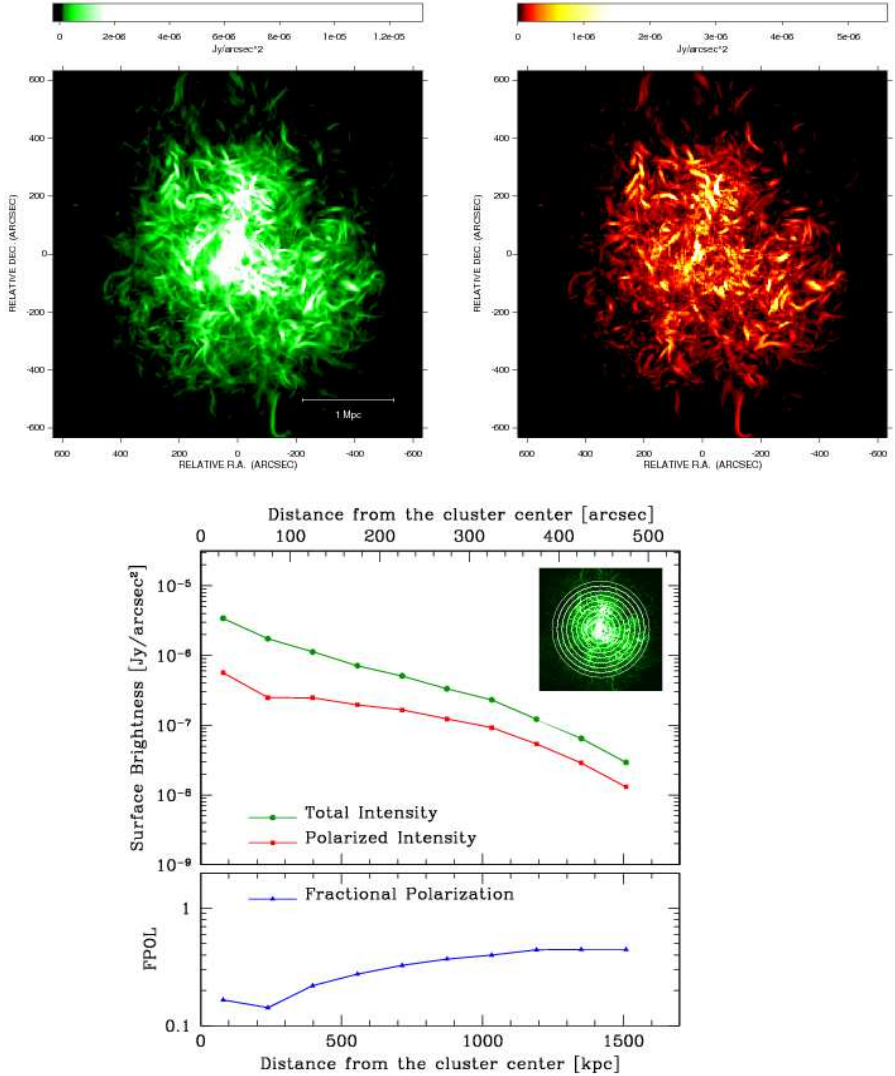}
\caption{Example of full resolution (10.7 kpc/pixel) radio halo 
in the simulated galaxy cluster R1a.
Left and right panels refer to the total intensity and polarized
surface brightness images, respectively.
The bottom panels show the azimuthally averaged radio-halo
brightness profiles of the total intensity $I$ (green dots), 
polarized intensity $P$ (red squares), and fractional polarization $FPOL$ (blue triangles). The profiles have been calculated in concentric annuli, 
as shown in the inset panel.
}
\label{Fig1}
\end{figure*}

The combined effects of the central magnetic field strength and the radial 
decline in the simulations performed by Xu et al. (2012)
can produce a volume average magnetic field that matches the 
equipartition magnetic field estimates for observed radio halos 
and, at the same time, is able to explain the observed values of Faraday 
rotation measures of radio sources in clusters. 

In this work, we use the magnetohydrodynamical cluster
simulations by Xu et al. (2011, 2012) to investigate the
polarized surface brightness of mock radio halos.
Following Murgia et al. (2004), we generate synthetic radio halo
images by illuminating the cosmological magnetic fields
with a population of relativistic electrons.
The synthetic images are calculated on a grid of $(512)^3$ points
with a spatial resolution of 10.7 kpc and a field of view of 5.48 Mpc.
At each point, on the computational grid,
we calculate the total intensity and the intrinsic linear
polarization emissivity at 1.4 GHz, by convolving the emission spectrum of 
a single relativistic electron with the particle energy distribution 
of an isotropic population of relativistic electrons: 
$N(\gamma,\theta)=K_{\rm 0}\gamma^{-\delta}(\sin\theta)/2 $,
where $\gamma$ is the electron's Lorentz factor, and $\theta$ is the pitch 
angle between the electron's velocity and the local direction of the 
magnetic field.
To be consistent with the radio halo total intensity images presented by
Xu et al. (2012), we adopt the same distribution for
the relativistic particles (whose parameters are summarized 
in Table~\ref{electrons}),
and we assume equipartition between the magnetic field ($u_B$)
and the relativistic electron ($u_{el}$) energy density at every
location in the intracluster medium.

\begin{figure*}
\centering
\includegraphics[width=16 cm]{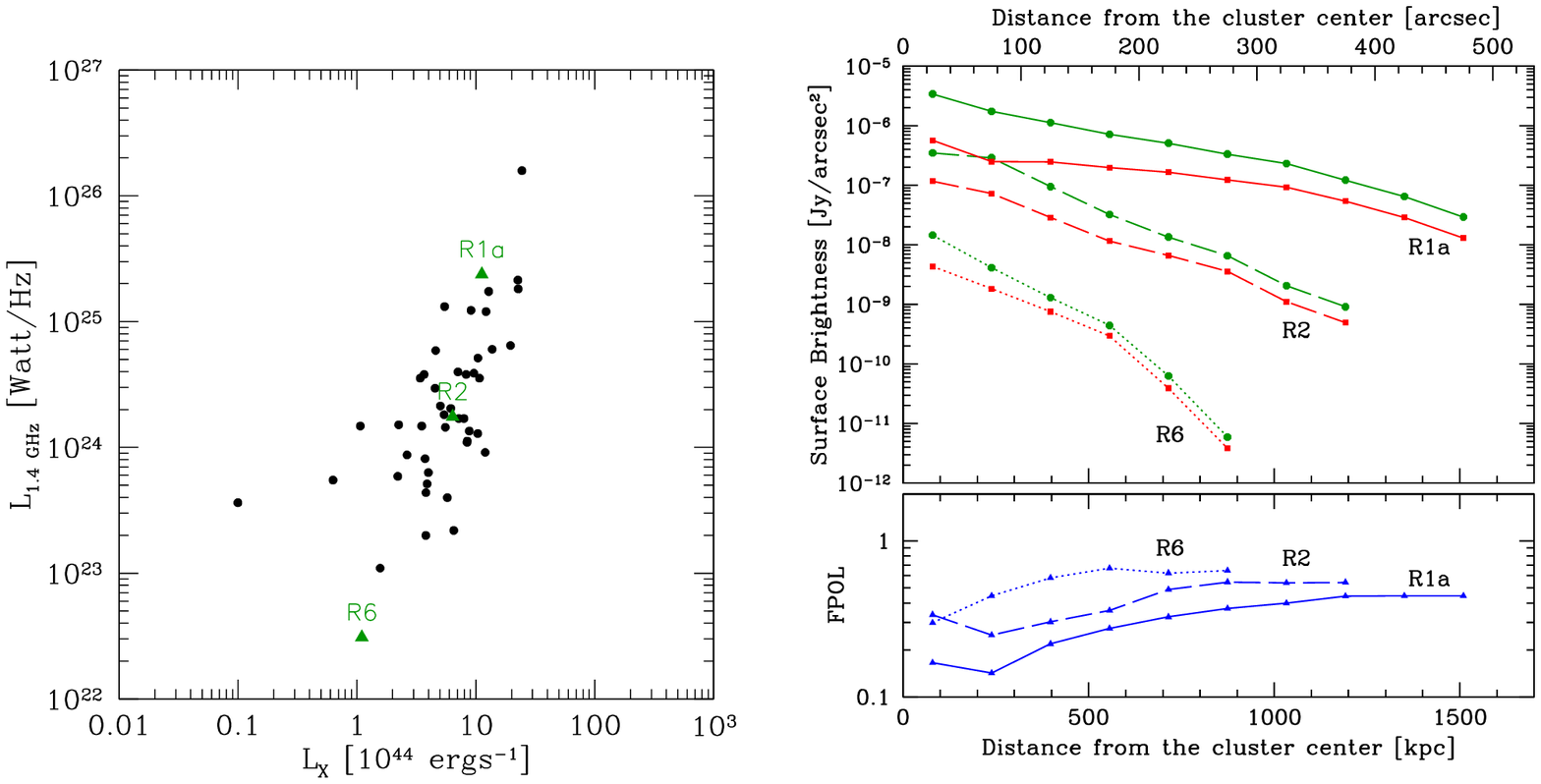}
\caption{Left: radio power at 1.4 GHz versus
the cluster X-ray luminosity in the 0.1$-$2.4 keV band.
Full dots are observed clusters taken from the recent compilations 
by Feretti et al. (2012) and Govoni et al. (2012);
green triangles are the simulated clusters R1a, R2, and R6 from Xu et al.
(2012).
In these simulations $L_X$ and $L_{\rm 1.4 GHz}$ have been calculated 
within a circle of 1 Mpc in radius.
Right: azimuthally averaged radio-halo
brightness profiles of the total intensity $I$ (green dots), 
polarized intensity $P$ (red squares), and
fractional polarization $FPOL=$ (blue triangles) of the 
clusters R1a (solid lines), R2 (dashed lines), and R6 (dotted lines).
}
\label{Fig2}
\end{figure*}

The Stokes parameters $Q$ and $U$, the polarized 
intensity $P=(Q^2+U^2)^{1/2}$, and polarization 
plane $\Psi=0.5\tan^{-1}(U/Q)$ images, have been calculated 
by taking into account that the polarization plane of the radio 
signal is subject to the  
Faraday rotation as it traverses the magnetized intracluster medium. 
Therefore, the integration of the polarized emissivity
along the line of sight has been performed 
as a vectorial sum in which the intrinsic polarization angle of the 
radiation coming from the simulation cells located at a depth $L$ is rotated 
by an amount: 

\begin{equation}
 \Delta\Psi = RM \times \left(\frac{c}{\nu}\right)^2, 
\label{rm1}
\end{equation}
where the rotation measure RM is given by
\begin{equation}
RM_{\rm~[rad/m^2]}=812\int_{0}^{L_{[kpc]}}n_{e~[cm^{-3}]}B_{\parallel~[\mu G]}dl.
\label{rm2}
\end{equation}
Here, $B_{\parallel}$ is the magnetic field along the line-of-sight.
This effect leads to the so-called internal depolarization 
of the radio signal.

We present the results for some simulated clusters
taken from the sample by Xu et al. (2011). In particular, we focus on
three clusters (labeled with R1a, R2, and R6) that are characterized 
by virial masses
ranging from $9.897\times10^{13}$ to $1.252\times10^{14}$ M$_{\odot}$ and
that have different final magnetic field strengths. Their properties are
summarized in Table 2.
These clusters have been obtained by performing a
self-consistent adaptive mesh refinement MHD simulation with initial 
magnetic fields injected by a single active galactic nucleus (AGN) 
using ENZO (Collins et al. 2010).

In Fig.~\ref{Fig1}, we show the results for the simulated cluster R1a,
which is among the most massive and most 
luminous systems in the simulated cluster sample by Xu et al. (2011).
The magnetic field strength at the cluster center is $\simeq$2.5$\mu$G. 
The full resolution images presented in Fig.~\ref{Fig1} 
have been performed at 1.4 GHz (with a bandwidth of 25 MHz) and are mapped 
as they would appear on the sky at a redshift\footnote{At this redshift 
and with the adopted cosmology, 1\arcsec~corresponds to 3.18 kpc.}  
z=0.2 (the current known average distance of the radio halos, 
Feretti et al. 2012).
The left panel refers to the total intensity $I$ image, 
while the right panel refers to the polarized intensity $P$ image.
It is not only possible to appreciate the full extension of the radio
halo emission both in total intensity and polarization
but also the fine details of its filamentary structure.
The total intensity surface brightness averaged in a circle of 500 kpc 
in radius was determined to be $<I>\simeq 1.51 \times 10^{-6}$ Jy/arcsec$^2$
while the polarization intensity surface brightness averaged 
in the same area was determined to be $<P>\simeq 2.75\times 10^{-7}$ Jy/arcsec$^2$.
Thus, the fractional polarization ($FPOL=P/I$)
for this simulation was found to be on average $<FPOL>\simeq 18$\%.

The bottom panel of Fig.~\ref{Fig1} shows the azimuthally 
averaged radio-halo brightness profiles of $I$, $P$, and $FPOL$.  
Each point represents the intensity, which is azimuthally averaged 
in concentric annuli 
of 50$''$ width centered on the cluster center.
The observed intensity profiles are traced up to a projected 
distance from the cluster 
center of 1.5 Mpc. At this distance from the cluster center, 
the fractional polarization is as high as 45\%.
The internal depolarization is stronger where the cluster RM is higher
and where the magnetic field is more disordered.
According to Eq.~\ref{rm2}, the central regions of a cluster, 
where the magnetic field strength and the gas density are higher, 
result in a higher RM. Therefore, the 
cluster center is expected to have a lower fractional polarization.
Indeed, the fractional polarization decreases 
down to $FPOL\simeq 15$\% near to the cluster center.

\begin{table*}
\caption{Properties of simulated galaxy clusters.}  
\label{simul}      
\centering          
\begin{tabular}{l c c c c c c}    
\hline\hline       
    Cluster   &   M$_{virial}$          & T    & n$_0$              & L$_X$  & B$_0$               & $L_{1.4 GHz}$\\  
              &   M$_{\odot}$           &  keV & $10^{-3}$cm$^{-3}$  &  erg/s  & $\mu$G            & Watt/Hz  \\
\hline   
     R1a      & $1.252 \times 10^{15}$  &7.65& 7.0  & $11.23\times10^{44}$ &  $\simeq$2.5   & $2.4\times10^{25}$ \\
     R2       & $8.633 \times 10^{14}$  &5.90& 5.6  & $6.39\times10^{44}$  &  $\simeq$1.0   & $1.8\times10^{24}$\\
     R6       & $9.897 \times 10^{13}$  &1.41& 10.3 & $1.17\times10^{44}$  &  $\simeq$0.5 & $3.1\times10^{22}$\\ \hline   
\multicolumn{7}{l}{\scriptsize Col.1: Galaxy cluster simulation; Col.2: Total virial mass; Col.3: Temperature;}\\
\multicolumn{7}{l}{\scriptsize Col.4: Central gas density; Col.5: Cluster X-ray luminosity in the 0.1-2.4 keV, which is}\\
\multicolumn{7}{l}{\scriptsize calculated within a circle of 1 Mpc in radius; Col.6: Central magnetic field strength;}\\
\multicolumn{7}{l}{\scriptsize Col.7: Radio power at 1.4 GHz, which is calculated within a circle of 1 Mpc in radius.}\\     
\end{tabular}
\end{table*}  

If we want to compare 
R1a with other radio halos that are known in the literature, it is important 
to note that this system ($L_{1.4GHz}=2.4\times10^{25}$ Watt/Hz), 
may represent the most luminous radio halos that are known so far. 
In the simulated cluster 
sample by Xu et al. (2011), other mock halos like R2 and R6 may be 
helpful in the comparison with fainter radio halos. 
The radio halo in R2 results in a power at 1.4 GHz of
$L_{1.4GHz}=1.8\times10^{24}$ Watt/Hz; thus, R2 represents 
intermediate power radio halos. 
Finally, the radio halo in the cluster R6 
($L_{1.4GHz}=3.1\times10^{22}$ Watt/Hz) is 
fainter than the halos that are known so far.
The radio power calculated at 1.4 GHz versus the cluster X-ray
luminosity are shown in the left panel of Fig.~\ref{Fig2}.
Full dots refer to the data published in the literature
(Feretti et al. 2012, Govoni et al. 2012, and references therein), 
while green triangles refer to the simulated clusters R1a, R2, R6. 
The synthetic radio halos lies perfectly in the relation between 
radio power versus X-ray luminosity ($L_{\rm 1.4 GHz}$ -- $L_{X}$). 
We note that in the $L_{\rm 1.4 GHz}$ -- $L_{X}$ relation a few observed 
galaxy clusters are over-luminous in radio with respect 
to the clusters X-ray luminosity. Xu et al. (2012) found that a simulated
cluster with multiple injection sources seems to fit in the relation
with the few outliers that are known in the literature.

The $I$, $P$, and $FPOL$ brightness profiles of R1a
are compared with those of R2 and R6 in the right panel 
of Fig.~\ref{Fig2}. 
At the center of R2, the $I$ and $P$ surface brightnesses
are about an order of magnitude fainter than in R1a, and the brightness 
profiles are steeper than in R1a. The $I$ and $P$  brightness profiles of
R6 are even fainter and steeper than in R2.
This behavior has been observed for real halos by Murgia et al. (2009).
In agreement with the fractional polarization profile of R1a,
the clusters R2 and R6 also show a smaller $FPOL$ 
close to the cluster center and a higher value at an increasing distance 
from the cluster centers.
The decrease of fractional polarization near the cluster 
center is due to internal depolarization.
Since the internal depolarization is strictly related to the cluster magnetic 
field strength, magnetic filed topology, and gas density, 
this trend is expected to change 
from cluster to cluster. We consistently found less 
internal depolarization in the clusters characterized by a lower magnetic 
field strength.

Although radio halos are affected
by internal depolarization, which is stronger where
the RM is higher (i.e. in more magnetized and denser
clusters) and where the magnetic field structure is more tangled, 
a significant polarization level is still expected as far 
as radio halos are observed at arbitrary high resolution 
and sensitivity.

\section{Polarization of cluster radio halos with current radio interferometers}

\begin{figure*}
\centering
\includegraphics[width=12 cm]{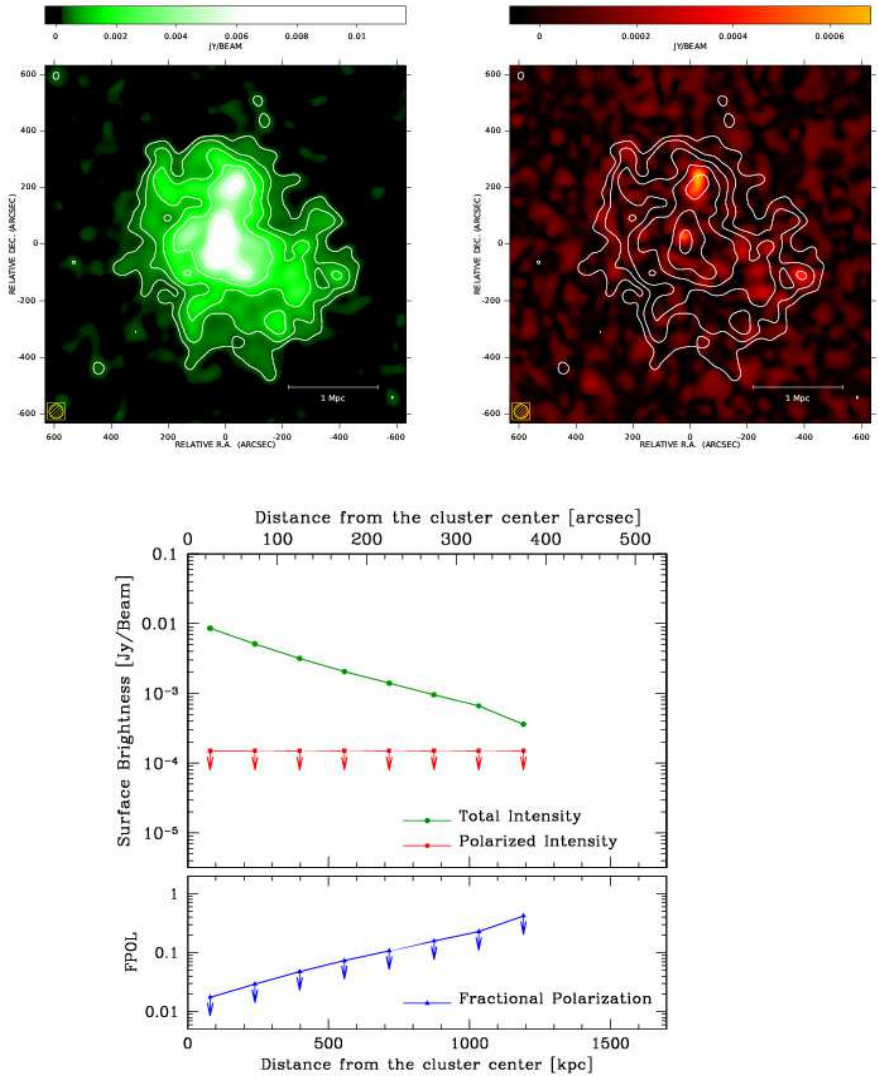}
\caption{Example of radio halo in the simulated 
galaxy cluster R1a.
Left and right panels refer to the total intensity and polarized
surface brightness images, respectively.
The images have been convolved to a resolution of 50$^{\prime \prime }$.
The white contour levels  
refer to the total intensity image at 50$^{\prime \prime }$ resolution.
Contour levels start at 0.3 mJy/beam (3-$\sigma$) and increase by 
a factor of 2.
The bottom panels show the azimuthally averaged radio-halo
brightness profiles of the total intensity $I$ (green dots), 
polarized intensity $P$ (red squares), and fractional polarization 
$FPOL$ (blue triangles). The upper limits are at 3-$\sigma$ level.
}
\label{Fig3}
\end{figure*}

The best images of radio halos available so far
have been obtained with sensitive interferometers like 
the Very Large Array (e.g. Clarke \& Ensslin 2006, Giovannini et al. 2009, 
Vacca et al. 2011), the Australia Telescope Compact Array (e.g. Liang et al. 2000), 
the Westerbork Synthesis Radio Telescope (e.g. Pizzo \& de Bruyn 2009, 
Brown \& Rudnick 2011, van Weeren et al. 2012), and the Giant 
Metrewave Radio Telescope 
(e.g. Venturi et al. 2007 and 2008, Bonafede et al. 2012). 
Experience shows that
observations with a resolution of $\sim$45-50$''$ are the best
suited to detecting and to imaging halo type sources.
Although many radio halos have been imaged well in total intensity, 
their polarized signals are still very difficult to be detected, and so far,
only a few radio halos have been imaged in polarization 
(Govoni et al. 2005, Pizzo et al. 2011, Bonafede et al. 2009).

We use the mock radio halos presented in Sect. 2
to investigate how instrumental effects in primis resolution and sensitivity
are related to the difficulties in detecting polarized emission from radio
halos. In particular, we investigate the polarized signal of 
the powerful radio halo R1a when observed with the sensitivity 
and resolution of the current radio interferometers.

\subsection{Sensitivity and resolution effects on radio halos}

A limited sensitivity imposes a threshold on the minimum detectable 
surface brightness, both in total intensity and in polarization.
At the same time, the low resolution needed to detect the faint radio halo
signal prevents the detection of the small-scale 
fluctuations of the surface brightness causing a suppression 
of the polarized intensity.
In fact, observing an extended source which is not uniformly 
polarized with a resolution larger than the angular scale
of coherent polarization patches leads to a decrease of the 
polarization signal.
The beam thus smoothes out the polarization of the source, 
and the measured polarization will be less than 
the true source polarization. This effect 
is called beam depolarization. 

To investigate how the radio halo R1a 
would appear when observed with a set-up typically used 
in many of the pointed interferometric observations of radio halos 
reported in the literature, we
convolved the full resolution $I$, $Q$, and $U$ images with a Gaussian 
beam of 50$^{\prime \prime }$, and we added a noise of
0.1 mJy/beam in the $I$ image and a noise of 0.05 mJy/beam in the 
$Q$ and $U$ images. These are typical resolution 
and sensitivity values achieved for a two-hour time-on-source 
observation with the VLA at 1.4 GHz in D configuration.
Finally, the convolved Q and U images were 
transformed back to $P$. The polarized intensity image $P$ 
has been corrected for the positive bias.

The resulting radio halo images in total intensity $I$ and
polarized intensity $P$ 
are shown on the left and on the right of Fig. ~\ref{Fig3}, respectively.
Most of the fine details of the halo 
structure, as seen in the full resolution images,
are not visible anymore. In total intensity, the radio halo appears 
smoother but still quite well detected. 
On the contrary, in polarized intensity, most of the halo emission 
falls below the noise level and the only 
few surviving localized spots are hardly detectable. 

In the bottom panels of Fig.~\ref{Fig3}, we show the azimuthally 
averaged radio-halo brightness profiles of $I$, $P$,
and $FPOL$.  
Each data point represents the average brightness in concentric annuli 
of a beam width centered on the cluster center.
The total intensity brightness profile is above
the 3-$\sigma_{I}$ up to about 1.2 Mpc from the cluster center, while the
polarized intensity falls below the 3-$\sigma_{P}$ noise level
everywhere along the profile.
Thus, the combination of a relatively shallow sensitivity and
the beam depolarization has a strong effect in determining
the observable polarization properties 
of radio halos. Even for a powerful radio halo like R1a
with a typical observing setup (two-hour of a VLA observation 
in D configuration), we do not expect to detect polarization
because of the current observational upper 
limits (e.g. Bacchi et al. 2003).

We checked that the radio halo is still visible for R2
(see also Xu et al. 2012) in total intensity but it is below the 
threshold of the minimum detectable polarization. 
R6 is not detectable either in total intensity or in polarization. 

Therefore, our simulations suggest that, although radio halos 
can be intrinsically polarized, detecting this polarized signal 
is a very hard task with the current radio facilities.
We need observations at a high resolution 
to reduce the beam depolarization of radio halos but, 
at the same time, at high resolution we need to increase the sensitivity 
to detect their low surface brightness.

\subsection{Resolution effects on radio halos}

In Sect. 3.1 we point out that the
sensitivity and the beam depolarization are two major effects 
that limit the possibility to detect the polarized signal from 
a radio halo. To understand the weight of these two limitations,
we focus on the effect of the beam alone, by neglecting the sensitivity 
of the observation.
We know the ``true'' 
polarization (see Sect. 2) of mock radio halos; thus, we can 
easily investigate how their fractional polarization decreases 
by increasing the observing beam resolution. 

To do this, we consider the cluster R1a and 
we convolve the full resolution $I$, $Q$, and $U$ images at 
different resolutions with a Gaussian beam of 10$''$, 30$''$, 50$''$, 
and 100$''$, without adding any noise. 
Finally, the convolved Q and U images are
transformed back to $P$.

\begin{figure}
\centering
\includegraphics[width=9 cm]{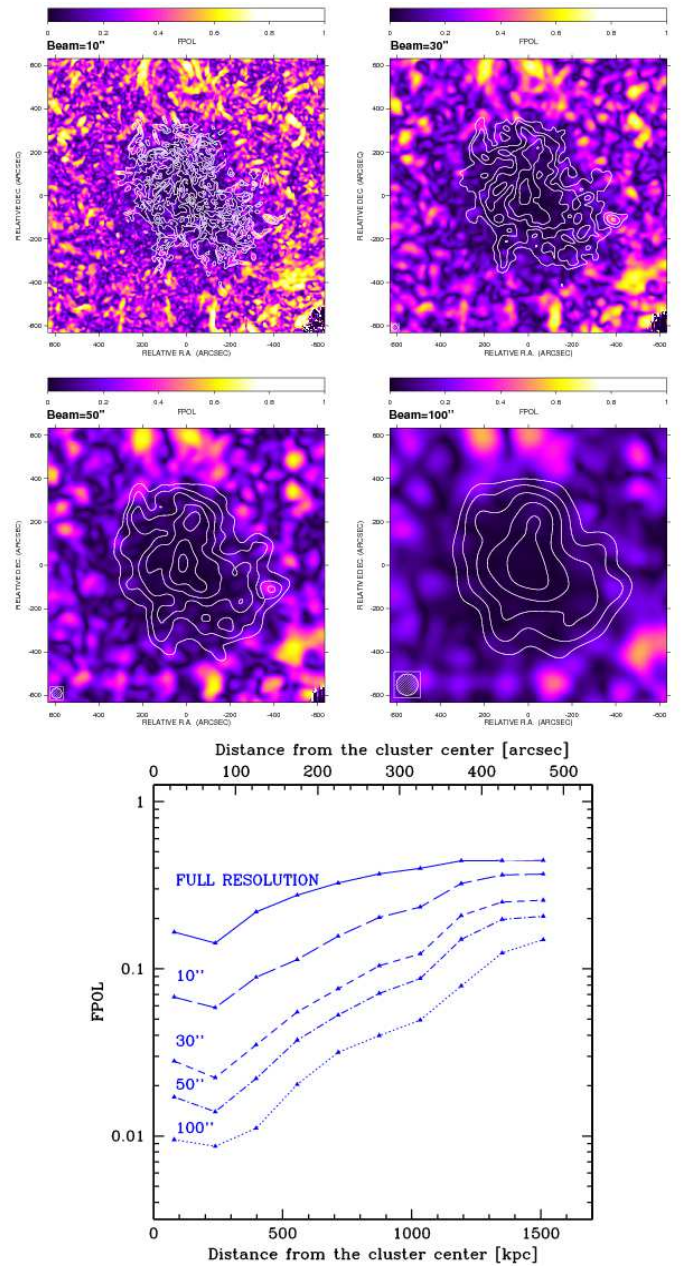}
\caption{Top: Fractional polarization ($FPOL=P/I$) images
of the radio halo, R1a, as expected at a resolution of 
10$''$, 30$''$, 50$''$, and 100$''$. 
Contour levels refer to the corresponding total intensity emission 
which are convolved at the same resolution.
Bottom: The azimuthally averaged radio-halo brightness profile
of the fractional polarization at full resolution (solid line)
is compared with the profiles at a resolution of 
10$''$ (long dashed line), 
30$''$ (short dashed line), 
50$''$ (dot dashed line), and 100$''$ (dotted line).
}
\label{Fig4}
\end{figure}

In the top panels of Fig.~\ref{Fig4}, we show the resulting 
fractional polarization images of the radio halo R1a, as expected 
at different resolutions.
In the bottom panel of Fig.~\ref{Fig4}, we show the azimuthally 
averaged radio-halo brightness profiles of the fractional polarization $FPOL$ 
images at different resolution.  
Each point represents the fractional polarization azimuthally averaged 
in concentric annuli of 50$''$ width centered on the cluster center.
The profiles at different resolutions are compared with the
``true'' polarization (at full resolution) shown in Fig.~\ref{Fig1}.
At a fixed distance from the cluster center, different observing beams
clearly affect the fractional polarization. In particular,
the fractional polarization for larger beams is drastically reduced. 
For example, at a distance of about 1.5 Mpc 
from the cluster center the fractional polarization 
of 45\% present at full resolution is reduced to about 36\% when 
the radio halo is observed with a beam 
of 10$''$ and reaches a value of 15\% if observed at a resolution of
100$''$.
The plot clearly shows that the beam depolarization also affects 
the trend of the polarized emission of radio halos as a function 
of the projected distance from the cluster center. 
By increasing the observing beam, the depolarization
is higher at the cluster center. This effect can be explained
by the fact that the polarization vectors are 
intrinsically more disordered near the cluster center 
than in the external part, and thus the beam depolarization is higher.
For example, the depolarization ($DP=FPOL_{int}/FPOL_{ext}$) between the internal 
($\simeq$100 kpc from the cluster center)
and the external ($\simeq$1.5 Mpc from the cluster center)
part of the halo in the full resolution 
image is $DP$$\simeq$0.37, but the fractional polarization is further 
reduced at the cluster center for lower resolution observation.
The depolarization is $DP$$\simeq$0.18 at a resolution of 10$''$, 
$DP$$\simeq$0.11 at 30$''$, $DP$$\simeq$0.08 at 50$''$,
and finally, $DP$$\simeq$0.06 at 100$''$ resolution.

It is important to note that processes related to merger 
events could change the magnetic field 
structure in galaxy clusters and may amplify the magnetic field strength. 
Both the magnetic field topology and the magnetic field strength
affect the fractional polarization in radio halos.
The observed trend of the fractional polarization versus the projected 
distance found for these mock radio halos can be used 
for comparison with future observations to constrain the cluster 
magnetic field properties. 
Indeed, both the strength and the morphology (i.e. the power spectrum) 
of cluster magnetic fields affect the polarization properties of radio
halos, and the consequent internal depolarization (Sect. 2) and
beam depolarization investigated here.
Thanks to these simulations, we are able to separate the contribution
of the intrinsic internal depolarization from that of the
extrinsic beam depolarization.

\section{Polarization of cluster radio halos with forthcoming 
radio interferometers}

Tight constraints on the magnetic field power 
spectrum of the galaxy clusters could be potentially obtained 
by detecting the radio halo polarization fluctuations and 
not just total intensity fluctuations (Vacca et al. 2010). 
In fact, the fractional polarization 
is a very robust indicator of the intracluster
magnetic field power spectrum because it only marginally depends 
on the energy spectrum of the synchrotron electrons and on 
the equipartition assumption.
Therefore, it would be very important to improve the sensitivity 
and angular resolution (as discussed in the Sect. 3)
of the future observations to detect polarized signal 
in as many radio halos as possible.

A new era of the study of the intracluster magnetic field will
begin with forthcoming radio interferometers. 
The high-sensitivity, low-instrumental polarization contribution, 
large-bandwidth, and large field of view of future instruments 
will allow high accuracy investigations of the intracluster magnetic fields 
in the central regions of the galaxy clusters and in the outskirts, 
where the magnetic fields are weaker.

On the basis of the cosmological magnetohydrodynamical simulations 
with initial magnetic fields that are injected by single active 
galactic nuclei (Xu et al. 2012), we  
explore the potential of the forthcoming
radio interferometers to detect the polarized emission
from radio halos.
In Sect. 4.1, we investigate how radio halos will appear
at 1.4 GHz when observed with a bandwidth of 300 MHz. 
Future sky surveys performed with SKA precursors and 
pathfinders are planned at this frequency and bandwidth.
In Sect. 4.2, we investigate pointed observations
performed at 1.4 GHz with a bandwidth of 1 GHz. This wide bandwidth
is provided for the SKA and is already available for the 
Jansky Very Large Array (JVLA).

If observations are carried out within a wide frequency band, 
the polarized vector may rotates by a large amount within the band,
and the polarized signal can be affected by a strong bandwidth depolarization.
The high sensitivity provided by the incoming radio instrumentation 
will make possible to observe the polarized signal in narrow 
frequency channels over large bandwidths, and hence to exploit
the RM synthesis (Brentjens \& de Bruyn 2005, Pizzo et al. 2011,
Macquart et al. 2012) at its best.
Pizzo et al. (2011) showed that this powerful technique is very effective
method to study the polarized emission of both the discrete sources 
and the diffuse emission in a comprehensive way.
In the simulations presented in the following, we apply the RM-synthesis 
to recover polarized signal reduced by the bandwidth depolarization. 
The details of the RM-synthesis procedure are given in a forthcoming paper 
(Murgia et al. in preparation).

\subsection{Sky surveys with SKA precursors and pathfinders}

In preparation for SKA, several next-generation radio telescopes 
and upgrades are being constructed around the world.
Among them, JVLA (USA), APERTIF (The Netherlands), 
ASKAP (Australia), and Meerkat (South Africa) are very good candidates 
to explore the polarization 
properties of cluster diffuse emission at GHz frequencies. 
Deep total intensity and polarization sky surveys have been planned 
for many of these telescopes and in the following we investigate the
potential of some of these surveys in 
detecting polarized emission from radio halos.

The survey WODAN (Westerbork Observations of the Deep APERTIF
Northern-Sky; R{\"o}ttgering et al. 2011) will use
APERTIF to explore the northern sky (declination $>$ $+30\deg$) 
at 1.4 GHz with a large bandwidth of 300 MHz. 
This survey will provide a spatial resolution of $\simeq$15$''$ and 
a sensitivity of about 10$\mu$Jy/beam.
Similar performances will be reached in the southern sky with ASKAP
through the total intensity survey EMU (Evolutionary Map 
of the Universe; Norris et al. 2011) and the 
polarization survey POSSUM 
(POlarization Sky Survey of the Universe's Magnetism).
These surveys, their scientific goals, and the technical challenges, which 
have been addressed to maximize the scientific results, have been 
recently described by Norris et al. (2012).
All these surveys will play an important role for
the study of non-thermal cluster physics.
In this context, Cassano et al. (2012) derive the expected number of radio 
halos at 1.4 GHz to explore the potential of the EMU and WODAN surveys.
By restricting the clusters to a redshift $z<0.6$, they show
that these surveys have the potential to detect up to 200 new radio halos.

We investigate here how the total intensity and the 
polarized emission of radio halos with different size 
and radio power will appear in radio observations, whose
resolution and sensitivity are in line with those expected for these
future sky surveys.

\begin{figure*}[h]
\centering
\includegraphics[width=14 cm]{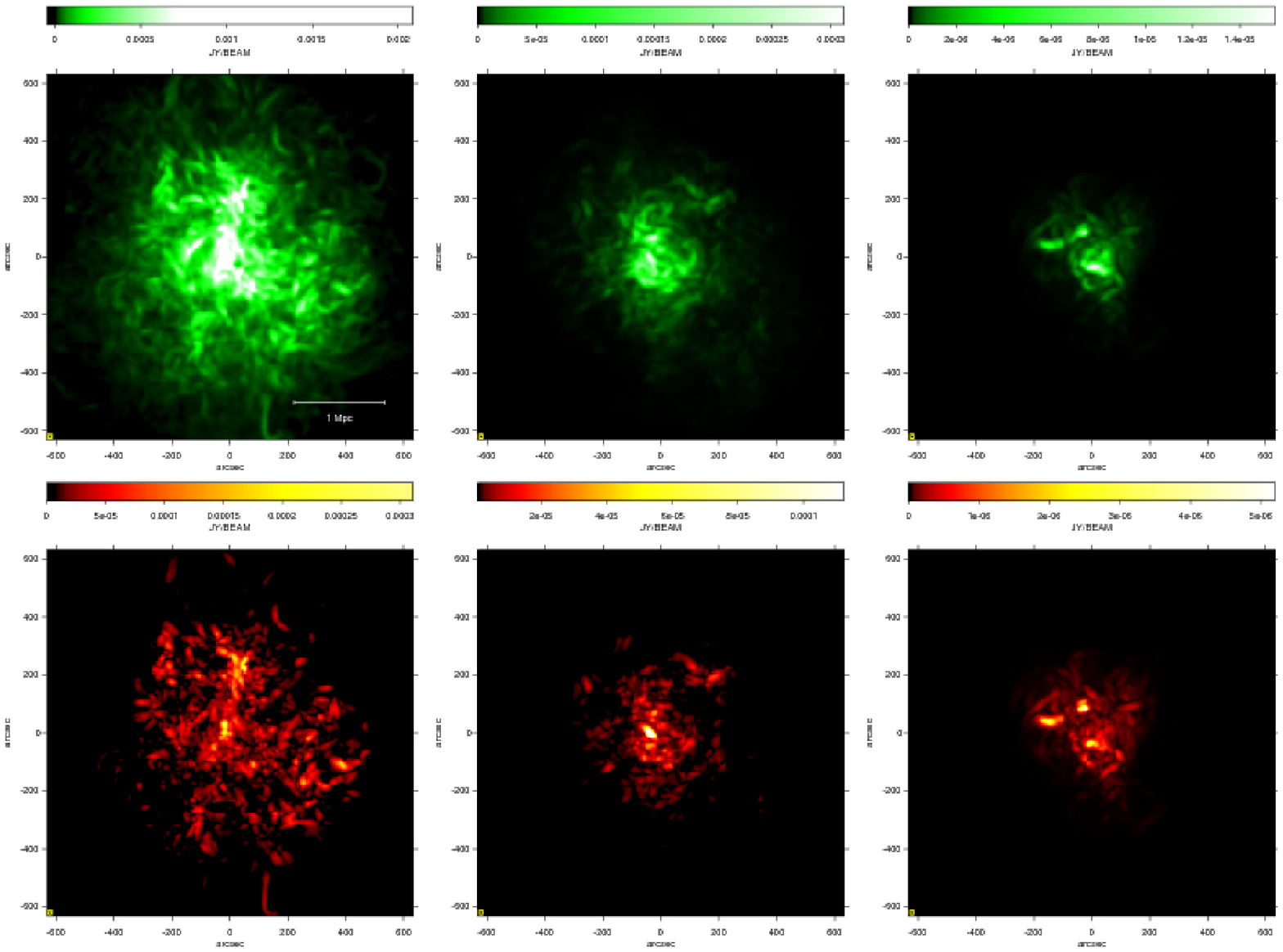}
\caption{
Mock radio halos taken from the simulations of R1a (left), R2 (middle) 
and R6 (right) by
Xu et al. (2012). 
Total intensity (top) and polarized (bottom)
surface brightness images are smoothed to a
resolution of 15$^{\prime \prime }$.}
\label{apertif1}
\end{figure*}

The top panels of Fig.~\ref{apertif1}
show the total intensity images of the simulated clusters R1a, R2, and R6
smoothed to a resolution of 15$^{\prime \prime }$.
The bottom panels of Fig.~\ref{apertif1}
show the polarized images of the R1a, R2, and R6
simulated clusters smoothed to a resolution of 15$^{\prime \prime }$, after that
the RM-synthesis is applied.
For each cluster, we produced Q and U image cubes that were 300 
channels by 1 MHz each, and the RM-synthesis was applied to the cubes.
This technique derotates the Q and U vectors for each pixel in each 
frequency channel image to compensate for a certain assumed 
rotation measure. An RM-synthesis cube was formed and the polarized signal 
was recovered by integrating over the Faraday depth space.

Fig.~\ref{apertif1} shows that
the fine details of the radio halos at this resolution 
are imaged well and the polarized surface 
brightness emission follow the
structure of the total intensity.
Accordingly with the different properties of the hosting clusters,
the three mock halos appear on the sky with
different surface brightness and angular extension.
R1a, the most massive and the higher magnetized, is the brightest
and the most extended of the three systems.  

\begin{figure*}[h]
\centering
\includegraphics[width=14 cm]{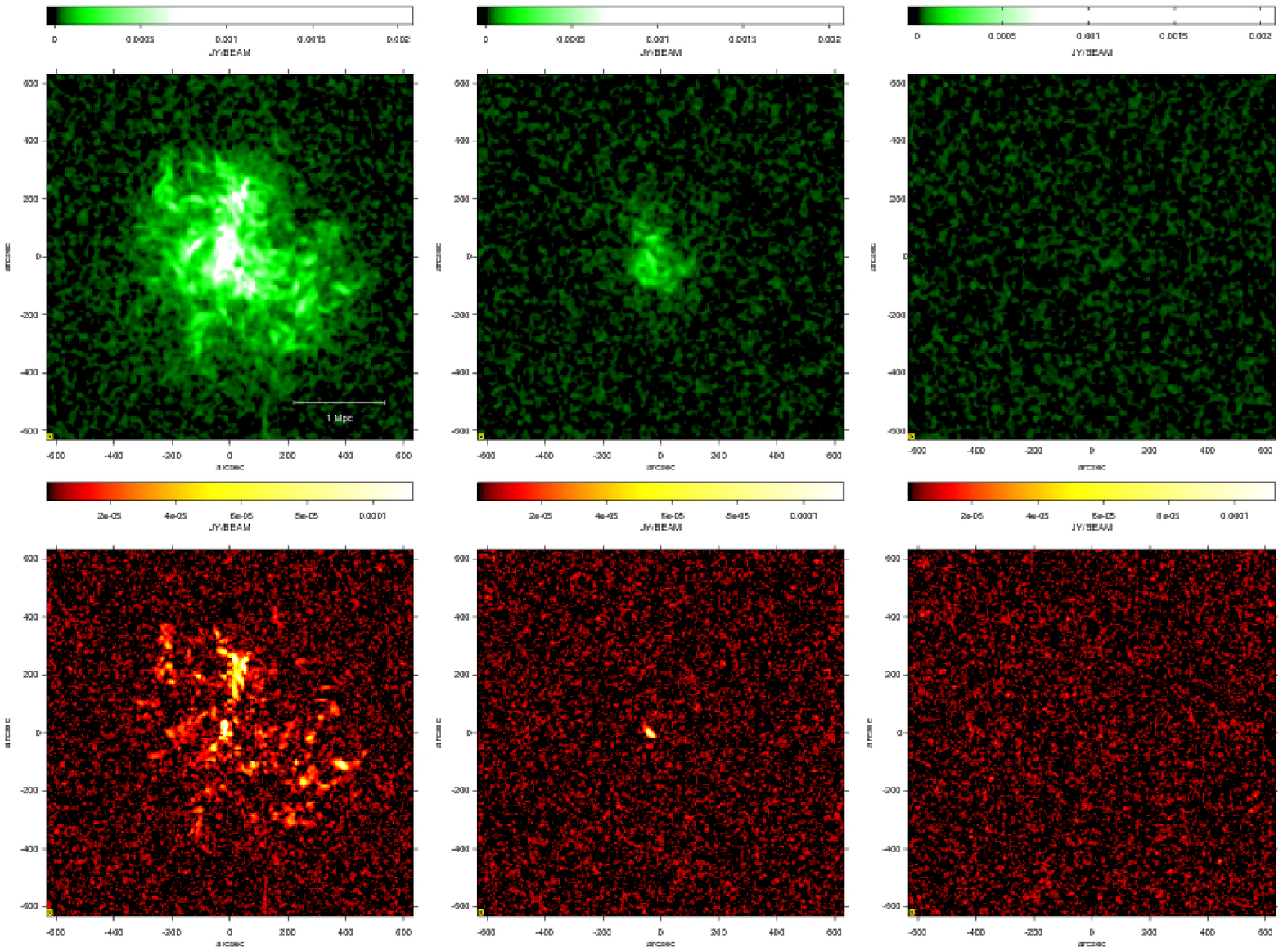}
\caption{
Mock radio halos taken from the simulations of R1a (left), R2 (middle) and R6
(right) by Xu et al. (2012). 
Total intensity (top) and polarized (bottom)
surface brightness images are shown as they
would appear when observed at 1.4 GHz with a bandwidth of 300 MHz, 
at a resolution of 15$^{\prime \prime }$, a noise level 
(1-$\sigma$) of 10$\mu$Jy/beam in I, and 5$\mu$Jy/beam in Q and U.
These resolutions and sensitivity values are in line with that 
expected in future sky surveys at 1.4 GHz performed with SKA precursors and 
pathfinders.}
\label{apertif2}
\end{figure*}

In agreement with observations (e.g Govoni et al. 2001,
Orr{\`u} et al. 2007, Murgia et al. 2009), the brightness profiles of
these mock radio halos smoothly decrease with distance from the cluster
center. Therefore, the outermost low brightness regions of the
halos become difficult to detect when the sensitivity limit is introduced 
and the entire radio halo may be missed
if its surface brightness is below the sensitivity threshold of 
the observation. 
Since the surface brightness is typically higher in powerful radio halos,
the detection of polarized emission will be easier in powerful radio halos.
In the top panels of Fig.~\ref{apertif2}, the same clusters are shown 
as they would appear when a noise level (1-$\sigma$) 
of 10$\mu$Jy/beam is considered.
This value is approximately the confusion limit at 
this resolution (Condon 1974).

In the bottom panels of Fig.~\ref{apertif2}, the same clusters are shown as they
would appear in polarization when a noise level is considered.
Determining the sensitivity in polarization is not easy since the 
expected polarization confusion limit is not obvious. 
For the Q and U images, we assume as a reference
a 1-$\sigma$ sensitivity value which is half the quoted value for 
the total intensity sensitivity (i.e. 5$\mu$Jy/beam in Q and U), 
being the sensitivity in 
polarization typically better than in total intensity.

Fig.~\ref{apertif2} clearly shows that this observing setup 
is very promising for detecting the polarized
emission of the most powerful ($>$$10^{25}$ Watt/Hz)
radio halos.
Therefore, surveys, like WODAN with APERTIF, and POSSUM with ASKAP,
may be useful possibilities to investigate 
the morphology of bright radio halos, both in total intensity and polarization, 
at 1.4 GHz. Indeed, our results suggest that the future surveys 
have enough sensitivity to detect the polarization emission 
up to the outermost low brightness regions of the most powerful radio
halos. On the other hand, radio halos of intermediate power
with $L_{1.4GHz}\simeq10^{24}$ Watt/Hz, 
like R2, will be imaged at high
resolution in total intensity, but their polarization will still be
hardly detectable.
Finally, the fainter radio halos, like R6, will definitely need a higher 
brightness sensitivity to be detected both in total intensity 
and in polarization.

\subsection{Pointed observations with SKA, its precursors, and its pathfinders}
One of the main difficulties in the study of radio halos is 
their faintness, which complicates their detection and complicates
their study with the current resources. A huge sensitivity to extended 
low-surface brightness emission is needed for a proper study of the radio 
halos both in total intensity and in polarization. 

The SKA will provide an unprecedented improvement in collecting area, thus
providing the necessary sensitivity to study the low-surface brightness  
radio halo emission at an angular resolution much higher than
the study previously done.
The simulations presented in this work offer the possibility of testing
whether these improvements are suitable for detecting the faint polarized
emission originating in the intracluster magnetic field 
filamentary structures.

\begin{figure*}[t]
\centering
\includegraphics[width=14 cm]{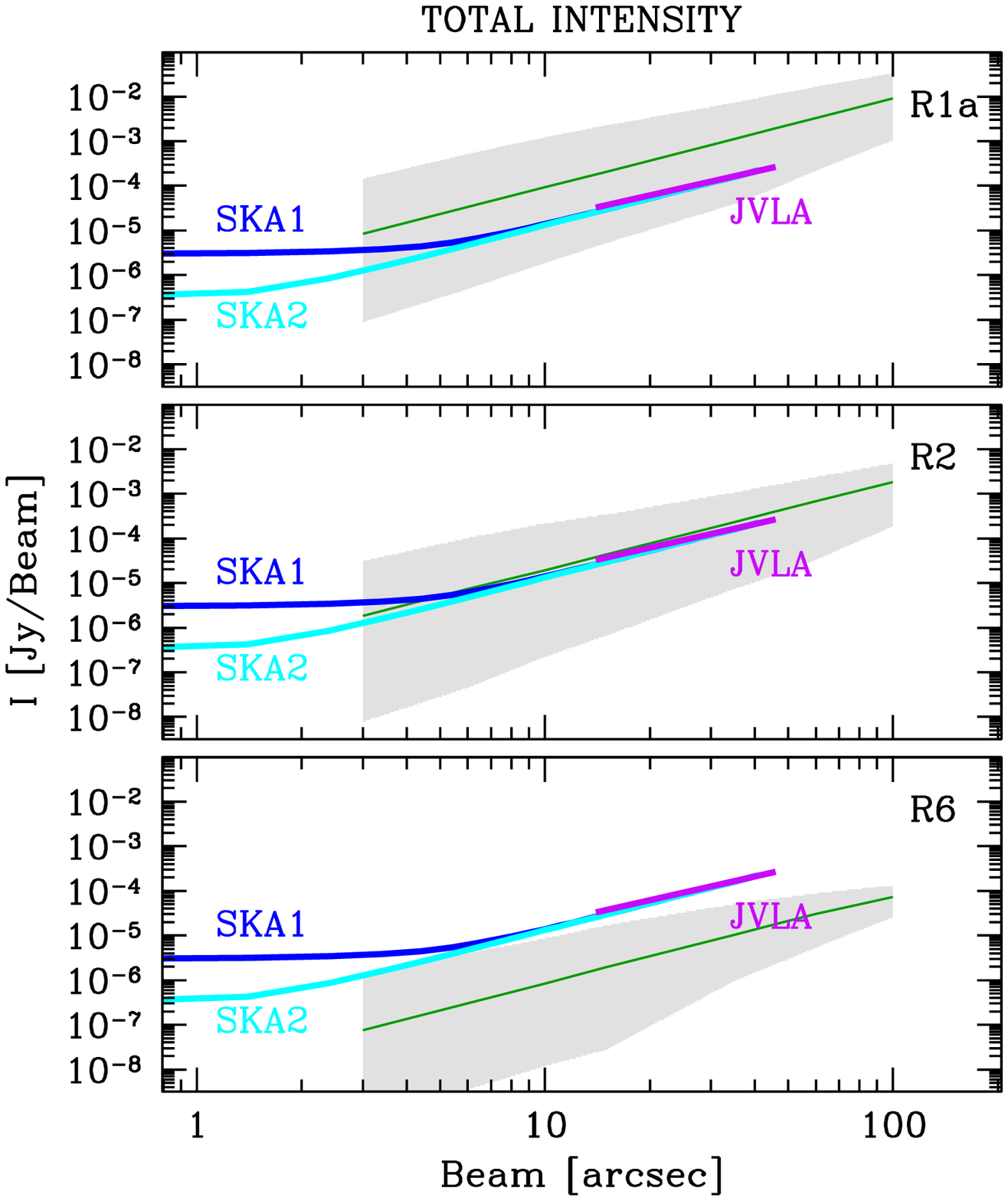}
\caption{
Total intensity surface brightness as a function of the beam size 
(3$''$-100$''$) for the mock radio halos 
R1a (top), R2 (middle), and R6 (bottom).
For each radio halo, the surface brightness is calculated within 
a circle of 3$r_e$ in radius. 
The solid thin green line shows the average brightness,
while the shaded region shows the maximum and minimum brightness 
fluctuations. 
The simulated surface brightness is compared with the sensitivity 
of wide band instruments (SKA Phase-1, SKA Phase-2, JVLA)
to explore their potential in detecting the total intensity 
emission of halos at different radio power.
The sensitivity refers to the 3-$\sigma$ limit.
}
\label{Fig7}
\end{figure*}

\begin{figure*}[t]
\centering
\includegraphics[width=14 cm]{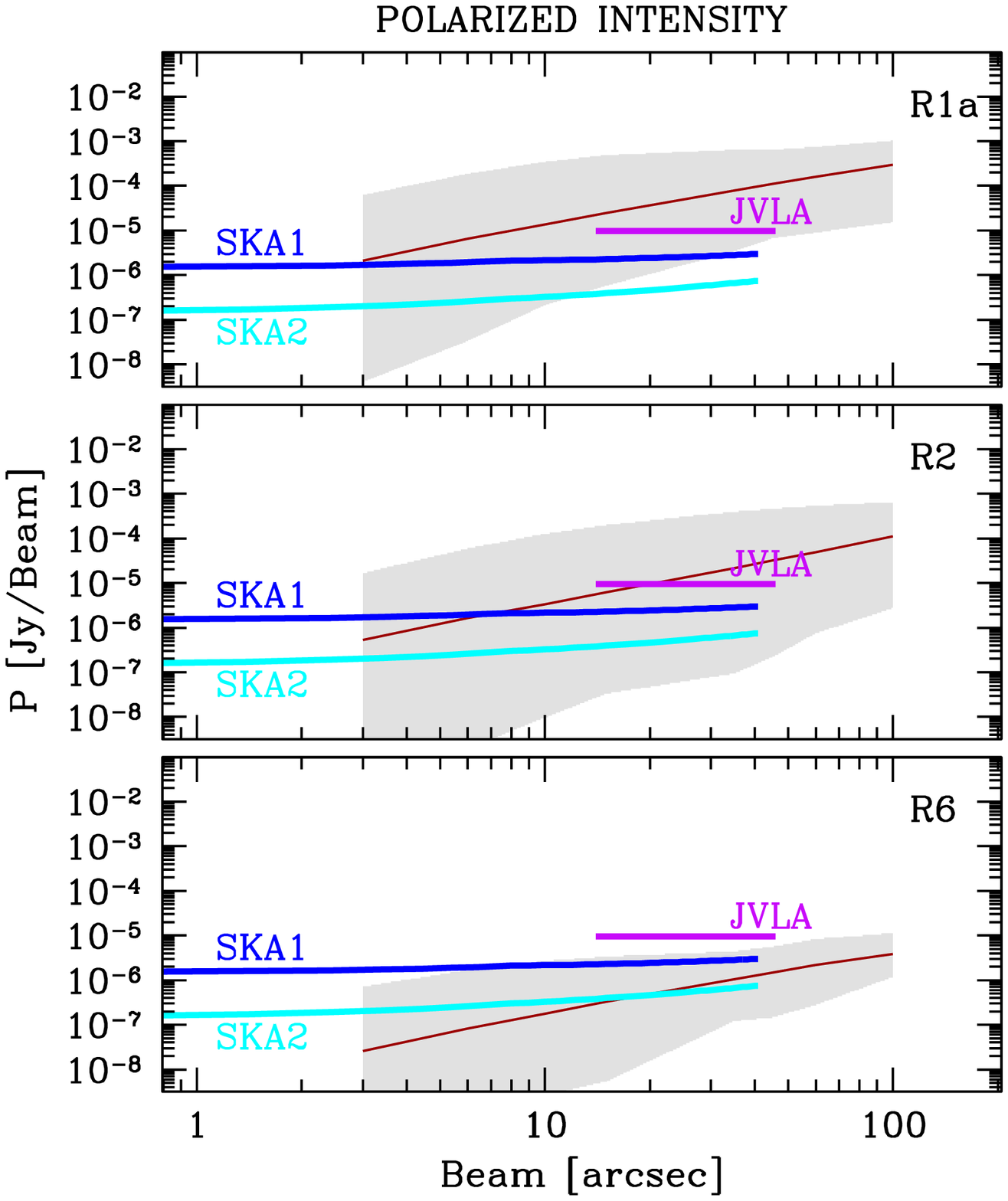}
\caption{
Polarized intensity surface brightness as a function of the beam size 
(3$''$-100$''$) for the mock radio halos 
R1a (top), R2 (middle), and R6 (bottom).
For each radio halo, the surface brightness is calculated within 
a circle of 3$r_e$ in radius. 
The solid thin red line shows the average brightness,
while the shaded region shows the maximum and minimum brightness 
fluctuations. 
The simulated surface brightness is compared with the sensitivity 
of wide band instruments (SKA Phase-1, SKA Phase-2, JVLA)
to explore their potential in detecting the polarized intensity 
emission of halos at different radio power.
The sensitivity refers to the 3-$\sigma$ limit.
}
\label{Fig8}
\end{figure*}

We consider how the total intensity and the polarized surface brightness 
of the mock halos of R1a, R2, and R6 would appear 
as seen at different resolutions.
In practice, we produce a set of images 
(similar to those shown in Fig.~\ref{apertif1}) at 1.4 GHz 
with a bandwidth of 1 GHz by varying the beam size 
in the range from 3$''$ (the full resolution
of our simulations) to 100$''$.
We then analyze the expected fluctuations of the radio brightness 
in the central region of the simulated halos and compare them
with the sensitivity expected for observations 
performed with SKA (see e.g. The Square Kilometer Array Design 
Reference Mission). In Figs.~\ref{Fig7}$-$~\ref{Fig8}, we show the results 
of this analysis.

Since we are comparing radio halos of different luminosities 
(see Fig.~\ref{Fig2}),
we need to rescale the region of interest according to the radio halo size.
As shown by Murgia et. al (2009), the e-folding radius r$_e$
of the surface brightness radial profile can be used as a reference for the 
radio halo size. Following the procedures described in 
Murgia et. al (2009), we obtained r$_e$=101\arcsec, 56\arcsec, 
and 30\arcsec~for R1a, R2, and R6, respectively by fitting the 
simulated radial profiles with an exponential.
Thus, we include in the statistics only those regions of the halo 
to within 3r$_e$ from the cluster center, for a proper comparison.
For each cluster, we plot the average brightness as a solid thin 
line and we use a shaded region to show the maximum and minimum brightness 
fluctuations 
calculated to within 3r$_e$ from the cluster center at that resolution.
The simulated surface brightnesses are compared with the sensitivity 
and the resolution of wide band instruments to explore
their potential in detecting the total intensity and polarized
emission of halos at different radio power.

In Fig.~\ref{Fig7}, we present the expected total intensity  
emission of the mock radio halos R1a (top), R2 (middle), and R6 (bottom) 
as a function of the angular resolution. 
We calculated, up to a resolution of $40''$,
the sensitivity achievable with SKA Phase-1 and SKA Phase-2 at 1.4 GHz 
for 1 hour of integration time and a bandwidth of 1 GHz.
The solid thick lines indicate the 3-$\sigma$ sensitivity limit
which is obtained by taking into account both the sensitivity reached for
an 1 hour of exposure time and the confusion limit. 
The sensitivity of the SKA Phase-2 at the few arcsec resolution
is an order of magnitude better than that of the SKA Phase-1. 
At a resolution higher than $\simeq$10$''$ the confusion limit 
is dominant, therefore, the sensitivity of the SKA 
Phase-1 and Phase-2 are the same.
We compare these limits with the expectations of the
JVLA, which represent the state-of-the-art instrument to investigate 
radio halos. In the resolution range between $15''$
to $45''$, the JVLA too is dominated by the confusion noise
for 1 hour of integration time. On the other hand,
the UV coverage of the JVLA does not permit to image the large 
angular structure of radio halos at higher resolution.

For a given resolution, we compare the average surface 
brightness of the mock radio halos R1a, R2, and R6  
with the minimum (3-$\sigma$) surface brightness that is detectable 
with the SKA (Phase-1 and Phase-2) and the JVLA.
Operationally, we consider that the halo emission is fully detectable 
only if the average surface brightness exceeds the
3-$\sigma$ sensitivity limit of the instruments. However, it is possible 
also to have partial detections if a few isolated patches 
have enough signal-to-noise to ``emerge'' above the sensitivity 
limit threshold.

Fig.~\ref{Fig7} shows that observations performed with the
SKA (Phase-1 and Phase-2) have the potential of detecting the 
total intensity emission in R1a-like and R2-like radio halos.
Similar results can also be obtained with the JVLA 
and with sky surveys that are planned with the SKA precursor and 
pathfinders (see Sect. 4.1), but the SKA will provide unprecedented
capabilities for imaging these radio halos at arcsecond resolution, 
where the confusion noise is negligible. 
On the other hand, SKA will miss the fainter R6-like halos.

In Fig.~\ref{Fig8}, we present the expected polarized
emission of the mock radio halos R1a (top), R2 (middle), and R6 (bottom)  
as a function of the angular resolution.
The analysis of polarization data performed with instruments
having a bandwidth of 1 GHz requires the RM-synthesis; therefore,
this technique is applied to the simulations in polarization.
We tentatively represent with solid thick lines a 3-$\sigma$ limit
expected with the SKA (Phase-1 and Phase-2) and the JVLA.
Determining the sensitivity in polarization is not easy, since the 
expected polarization confusion limit is not obvious. 
Therefore, we assume that the confusion limit is negligible in 
polarization, and we consider a putative polarization sensitivity
calculated as half the sensitivity expected in total intensity.
Under the above assumptions, the SKA Phase-2 will improve the
sensitivity to about one order of magnitude with respect 
the SKA Phase-1 and to about two orders of magnitude with respect to
the JVLA. 

Fig.~\ref{Fig8} shows that the sensitivity reachable with the SKA will  
detect the polarized emission in radio halos of high (R1a-like)
and intermediate (R2-like) luminosity.
Therefore, the detection of high-resolution polarized emission in 
radio halos with a luminosity of $L_{1.4GHz}=1-2\times10^{24}$ Watt/Hz 
requires a sensitivity reachable only with the SKA.
For fainter halos (R6-like), the partial detection of the 
polarized emission could be also possible, 
albeit very difficult even for the SKA Phase-2.
It is interesting to note that on the basis of our modeling,
the JVLA can already detect polarized emission from strong radio halos
at a relatively low resolution. Indeed, this is a
prediction of what we will be able to test in the near future
because of the wide-band polarized observations scheduled with the JVLA
at 1.4 GHz for a sample of radio halos. 

As a last point, we briefly address the scatter of the radio halo brightness
at a given resolution. The scatter of the simulated surface brightness
represented by the shaded region is very interesting in its own.
This intrinsic variation is due to two distinct effects. 
One is the general dimming of the radio halo surface brightness
with its radius and the other is the point-to-point scatter due 
to the filamentary structure of the intracluster magnetic field. 
The brightness dispersion increases by increasing the resolution
because the filamentary structure of the halo is progressively resolved. 
Therefore, the SKA will not only be able to detect the radio halo emission
but also to measure the intrinsic scatter of both the total intensity and the polarized surface brightness. Determining the amplitude of this scatter 
is important, since this quantity 
is strictly related to the power spectrum of the 
intracluster magnetic field fluctuations (Murgia et al. 2004, Govoni et 
al. 2006, Vacca et al. 2010).

\section{Conclusion}

In this work, we use the cosmological
magnetohydrodynamical simulations by Xu et al. (2012)
to predict the expected surface brightness distribution
of radio halos both in total intensity and in polarization.
Under the equipartition assumption and a reasonable shape
for the relativistic electron energy spectrum, 
we produce synthetic radio halo images at a frequency of 
1.4\,GHz for three simulated halos of representative total radio 
luminosity.

We first show that, simulated radio halos are intrinsically polarized
at full-resolution.
The fractional polarization at the cluster center
is $\simeq 15-35$ \% with values varying 
from cluster to cluster and increasing with the distance from
the cluster center. However, the polarized signal is undetectable
if observed with the comparatively shallow sensitivity and
low resolution of current radio interferometers.

We then use our modeling to investigate the potential 
of the increased sensitivity and the resolution of the forthcoming 
large radio telescopes.
We find that surveys planned with the SKA precursors 
will be in principle able to detect
the polarized emission in the most luminous halos known, while the
halos of intermediate and faint luminosity will still be hardly detectable.
Furthermore, we find that the JVLA have the potential to 
already detect polarized emission from strong radio halos.

For radio halos of strong and intermediate luminosity, 
we expect a polarized signal of about $2-0.5$ $\mu$Jy/beam at 1.4 GHz with
a resolution of 3$''$. The SKA, the most ambitious radio telescope
ever planned, could have the sufficient sensitivity to fully detect
the polarized emission of these radio halos at high resolution.
For fainter halos, the partial detection of the polarized emission 
could also be possible, albeit very difficult even for the SKA.

Furthermore, the SKA will be capable of measuring
the intrinsic scatter of both the total intensity and the polarized 
surface brightness fluctuations. 
Determining the amplitude of this scatter is important, 
since this quantity is strictly related to the power spectrum 
of the intracluster magnetic field fluctuations 
(Murgia et al. 2004, Govoni et al. 2006, Vacca et al. 2010).

\begin{acknowledgements}
 The authors thank Joseph Lazio for useful discussions. 
 This research was supported by PRIN-INAF2009 and by the
 project L.R. 7 Agosto 2007, N.7 : 
 "Promozione della Ricerca Scientifica e dell'innovazione 
 Tecnologica in Sardegna", Progetti di ricerca fondamentale o 
 di base annualit\'a  2012.
  H.X. and H.L. were supported by the LDRD and IGPP programs at LANL and by
the DOE/Office of Fusion Energy Science through CMSO. 
  Computing resources for cluster simulations were supplied by LANL on Institutional Computing Resource. 
  ENZO is developed at the Laboratory for Computational Astrophysics, UCSD, with partial support from NSF grants AST-0708960 and AST-0808184 to M.L.N.
\end{acknowledgements}


\begin{thebibliography}{}

\bibitem{} 
Bacchi M., Feretti L., Giovannini G., Govoni, F., 2003, A\&A, 400, 465

\bibitem{} 
Bonafede, A., Feretti, L, Giovannini, G., et al., 2009, A\&A, 503, 707 

\bibitem{} 
Bonafede, A., Feretti, L., Murgia, M., et al., 2010, A\&A, 513, A30 

\bibitem{} 
Bonafede, A., Govoni, F., Feretti, L., et al., 2011a, A\&A, 530, A24 

\bibitem{}
Bonafede, A., Dolag, K., Stasyszyn, F., Murante, G., Borgani, S., 2011b, \mnras, 418, 2234

\bibitem{} 
Bonafede, A., Br{\"u}ggen, M., van Weeren, R., et al., 2012, \mnras, 426, 40 

\bibitem{} 
Brentjens, M.~A., de Bruyn, A.~G., 2005, \aap, 441, 1217 

\bibitem{} 
Brown, S., \& Rudnick, L., 2011, \mnras, 412, 2 

\bibitem{} 
Br{\"u}ggen, M., Ruszkowski, M., Simionescu, A., Hoeft, M., Dalla Vecchia, C., 
2005, \apjl, 631, L21 

\bibitem{} 
Cassano, R., Brunetti, G., Norris, R.~P., et al.\ 2012, \aap, 548, A100 

\bibitem{} 
Clarke, T.E., \& Ensslin, T.A.\ 2006, \aj, 131, 2900 

\bibitem{}
Collins D. C., Xu, H., Norman, M. L., et al., 2010, \apjs, 186, 308

\bibitem{} 
Condon, J.~J.\ 1974, \apj, 188, 279 

\bibitem{} 
Dolag, K., Bartelmann, M., Lesch, H., 1999, \aap, 348, 351 

\bibitem{}
Dolag, K., Bartelmann, M., Lesch, H., 2002, \aap, 387, 383

\bibitem{} 
Dolag, K., Grasso, D., Springel, V., Tkachev, I., 2005, \jcap, 1, 9 

\bibitem{}
Dolag, K., Bykov, A.~M., Diaferio, A., 2008, \ssr, 134, 311

\bibitem{}
Donnert, J., Dolag, K., Lesch, H., M{\"u}ller, E. 2009, \mnras, 392, 1008

\bibitem{} 
Donnert, J., Dolag, K., Brunetti, G., Cassano, R., Bonafede, A.,
2010, \mnras, 401, 47 

\bibitem{}
Dubois, Y., \& Teyssier, R., 2008, \aap, 482, L13

\bibitem{}
Dubois, Y., Devriendt, J., Slyz, A., Silk, J., 2009, \mnras, 399, L49

\bibitem{} 
Feretti, L., Giovannini, G., Govoni, F., Murgia, M., 2012, \aapr, 20, 54 

\bibitem{} 
Giovannini, G., Bonafede, A., Feretti, L., et al., 2009, A\&A, 507, 1257  

\bibitem{} 
Govoni, F., En{\ss}lin, T.~A., Feretti, L., Giovannini, G., 2001, \aap, 369, 441

\bibitem{} 
Govoni, F., Murgia, M., Feretti, L., et al., 2005, \aap, 430, L5 

\bibitem{} 
Govoni, F., Murgia, M., Feretti, L., et al., 2006, \aap, 460, 425 

\bibitem{} 
Govoni, F., Ferrari, C., Feretti, L., et al., 2012, \aap, 545, A74 

\bibitem{} 
Liang, H., Hunstead, 
R.~W., Birkinshaw, M., \& Andreani, P., 2000, \apj, 544, 686 
 
\bibitem{} 
Macquart, J.-P., Ekers, R.~D., Feain, I., Johnston-Hollitt, M.,
2012, \apj, 750, 139 

\bibitem{} 
Murgia, M., Govoni, F., Feretti, L., et al., 2004, \aap, 424, 429 

\bibitem{} 
Murgia, M., Govoni, F., Markevitch, M., et al., 2009, \aap, 499, 679 

\bibitem{} 
Norris, R.P., Hopkins, A.M., Afonso, J., et al., 2011, \pasa, 28, 215 

\bibitem{} 
Norris, R.P., Afonso, J., Bacon, D., et al., 2012, accepted by PASA, arXiv:1210.7521 

\bibitem{} 
Orr{\`u}, E., Murgia, M., Feretti, L., et al., 2007, \aap, 467, 943 

\bibitem{} 
Pizzo, R.F., \& de Bruyn, A.~G., 2009, \aap, 507, 639 

\bibitem{} 
Pizzo, R.F., de Bruyn, A.G., Bernardi, G., Brentjens, M.A., 2011, A\&A, 525, A104 

\bibitem{} 
Roettiger, K., Stone, J.M., Burns, J.O., 1999, \apj, 518, 594 

\bibitem{} 
R{\"o}ttgering, 
H., Afonso, J., Barthel, P., et al.\ 2011, Journal of Astrophysics and 
Astronomy, 32, 557 

\bibitem{} 
SKA Science Working Group, 2011, The Square Kilometer Array Design 
Reference Mission: SKA Phase 1

\bibitem{} 
Tribble, P.C., 1991, \mnras, 253, 147 

\bibitem{} 
van Weeren, R.~J., Bonafede, A., Ebeling, H., et al., 2012, \mnras, 425, L36 

\bibitem{} 
Vacca, V., Murgia, M., Govoni, F., et al., 2010, \aap, 514, A71

\bibitem{} 
Vacca, V., Govoni, F., Murgia, M., et al., 2011, \aap, 535, A82 

\bibitem{} 
Venturi, T., Giacintucci, S., Brunetti, G., et al., 2007, \aap, 463, 937 

\bibitem{} 
Venturi, T., Giacintucci, S., Dallacasa, D., et al., 2008, \aap, 484, 327 

\bibitem{}
Widrow, L.M. 2002, Reviews of Modern Physics, 74, 775

\bibitem{}
Xu, H., Li, H., Collins, D.C., Li, S., Norman, M.L., 2009, \apjl,
  698, L14

\bibitem{} 
Xu, H., Li, H., Collins, 
D.C., Li, S., Norman, M.L., 2010, \apj, 725, 2152 

\bibitem{} 
Xu, H., Li, H., Collins, 
D.C., Li, S., Norman, M.L., 2011, \apj, 739, 77 

\bibitem{} 
Xu, H., Govoni, F., Murgia, M., et al., 2012, \apj, 759, 40 

\end{thebibliography}
\end{document}